\documentclass[12pt]{article}

\usepackage{amsmath,amssymb,epsfig}
\usepackage{graphicx}

\newcommand{\GeV}{\; {\mathrm{GeV}}}
\newcommand{\MeV}{\; {\mathrm{MeV}}}

\newcommand{\PPP}{P_{33}(1232)}
\newcommand{\PP}{P_{11}(1440)}
\newcommand{\DD}{D_{13}(1520)}
\renewcommand{\SS}{S_{11}(1535)}

\begin{document}
\setlength{\parskip}{0.45cm}
\setlength{\baselineskip}{0.75cm}

%
%
%
\begin{titlepage}
\setlength{\parskip}{0.25cm}
\setlength{\baselineskip}{0.25cm}
\vspace{1.0cm}
\begin{center}
\Large
{\bf Quark-Hadron Duality in Neutrino Scattering}

\vspace{1.5cm}

\large
O.~Lalakulich$^a$,
W.~Melnitchouk$^b$, and
E.~A.~Paschos$^a$   \\
\vspace{1.0cm}

\normalsize
$^a$
{\it Universit\"{a}t Dortmund, Institut f\"{u}r Physik,}\\
{\it D-44221 Dortmund, Germany} \\
\vspace{0.5cm}

$^b$
{\it Jefferson Lab, 12000 Jefferson Avenue}\\
{\it Newport News, VA 23606, USA}\\
\vspace{0.2cm}

\vspace{1.5cm}
\end{center}

\begin{abstract}
We present a phenomenological model of the quark-hadron transition in
neutrino-nucleon scattering.
Using recently extracted weak nucleon transition form factors, we
investigate the extent to which local and global quark-hadron duality
is applicable in the neutrino $F_1$, $F_2$ and $F_3$ structure
functions, and contrast this with duality in electron scattering.
Our findings suggest that duality works relatively well for
neutrino--nucleon scattering for the $F_2$ and $F_3$ structure
functions, but not as well for $F_1$.
We also calculate the quasi-elastic, resonance and deep inelastic
contributions to the Adler sum rule, and find it to be satisfied to
within 10\% for $0.5 \lesssim Q^2 \lesssim 2\GeV^2$.
\end{abstract}
\end{titlepage}

\section{Introduction}

Historically, neutrino scattering has provided vital information on the
structure of the nucleon, complementary to that obtained by the more
ubiquitous electromagnetic probes.
In deep inelastic scattering (DIS), neutrino-induced structure
functions have been used, in conjunction with electromagnetic structure
functions, as the primary tool to separate valence and sea quark
distributions.
Neutrinos are also necessary to complete our knowledge of the full
vector and axial vector structure of the nucleon elastic and transition
form factors.

At the parton level, deep inelastic structure functions describe
incoherent scattering of a hard probe from quarks and gluons
(generically, partons) in the nucleon; form factors, in contrast,
characterize the coherent or bound-state response of the nucleon
to an electromagnetic or weak probe.
While on the face of it the physics of coherent and incoherent
processes is rather distinct, they are in fact intimately related
through the phenomenon of quark-hadron duality.

Quark-hadron duality in structure functions refers to the observation,
first made by Bloom and Gilman \cite{Blo70}, that the average over
resonances produced in inclusive $eN$ scattering closely resembles the
leading twist (or ``scaling'') function measured in the deep inelastic
region.
Furthermore, as $Q^2$ increases the average over resonances approaches
the asymptotic scaling function.
Within QCD, the degree to which this ``Bloom--Gilman duality'' holds is
a direct reflection of the size of higher twist effects in the nucleon
\cite{DeR77,Ji95}.
According to the operator product expansion (OPE), higher twists are
related to nucleon matrix elements of multi-quark or quark-gluon
operators, which contain information on long-range, nonperturbative
interactions between partons.
Such interactions characterize the structure of the resonances and
diminish with powers of $1/Q^2$ as $Q^2\to\infty$.

Recently there has been a resurgence of interest in duality in electron
scattering at Jefferson Lab and elsewhere, where its target, flavor, spin
and nuclear dependence has been explored
\cite{Nic00,Chr02,Arr03,Liang04,Wesselmann:2006mw,Bur01,JLa01,Air03}.
Duality has been confirmed to good accuracy for the proton $F_2$ and $F_L$
structure functions to $Q^2$ values as low as 1~GeV$^2$ or even lower. The
basic features of duality have also been studied in terms of dynamical
models \cite{Clo01,Clo03,Isg01}, in phenomenological parametrizations of
the form factors \cite{Dav02,Mel01,Ste04}, as well as in the Rein-Sehgal
model \cite{Gra05} and (in the $\Delta$-resonance region) the Sato-Lee
model \cite{Mat05}.

Within the models, neutrino scattering can provide an important
consistency check, and lead to a better understanding of the
systematics of nucleon $N$ $\to$ resonance $R$ transitions.
While the phenomenological information on duality from electron
scattering has been steadily accumulating \cite{Mel05}, there is
at present almost nothing known empirically about the workings of
duality in neutrino scattering.
There are plans, however, to measure neutrino cross sections using
a high-intensity neutrino beam at Fermilab \cite{Dra04}.

In a parallel development, recent theoretical work has investigated
the excitation of resonances by neutrinos for both $J=3/2$
\cite{Pas03,Lal05} and $J=1/2$ resonances \cite{Lal06}.
In the latter work the weak vector form factors were determined from
Jefferson Lab data using the conserved vector current (CVC) hypothesis,
and two of the axial form factors from PCAC, although the $Q^2$
dependence was not very well constrained.
To date there are only rudimentary data on neutrinoproduction of
resonances beyond the $P_{33}(1232)$ region, however, more accurate
data are expected, and a precise comparison will be possible in the
future.
In this paper we use the recent theoretical results to perform a
detailed phenomenological study of duality in neutrino scattering.

If one assumes that duality holds for neutrino scattering, then the
average area under the resonances must follow the scaling curve.
In this case the results of our comparison can be interpreted as a
check on how well the $Q^2$ dependence of the transition amplitudes
$N \to R$ is known.
Deviations from duality would in this case provide information on
the size of the background, and of the axial form factors, which
were not determined in the model \cite{Lal06} (for example, the normalization
of $C_3^A$ and $C_4^A$ for the $D_{13}(1520)$ resonance, as well
as the $Q^2$ dependence of all axial form factors).
Obtaining a better understanding of the dynamics in this kinematic
region is also crucial for the interpretation of neutrino--oscillation
experiments \cite{Hay04}.

In Section~\ref{Formalism} we review the formalism which is used in
this study, and provide details about the transition form factor
parameterizations.
Results on local and global aspects of duality in neutrino scattering
are discussed in Sec.~\ref{LocalDuality}, and contrasted with duality
in electron scattering. We also discuss the saturation of the Adler Sum rule, including its contributions from resonances, quasielastic and DIS regions. Finally, in Sec.~\ref{Conclusions} we summarize our results and draw conclusions from our study.

\section{Formalism \label{Formalism}}

Testing the degree to which duality in lepton-nucleon scattering is
valid requires knowledge both of structure functions in the resonance
region, and of the scaling functions applicable in the DIS regime.
The former are calculated in terms of the nucleon $\to$ resonance
transition form factors, while the latter can be evaluated from
twist-two parton distribution functions.
In this section we review both of these inputs, firstly outlining
the parameterizations of the $N \to R$ transition form factors from
which the resonance structure functions are computed, and then
summarizing the essential formulas for the twist-two structure
functions.
A more complete account of the formalism can be found in
Refs.~\cite{Lal05,Lal06}; here we shall present only those details
which are pertinent to the specific discussion of duality.

\subsection{Weak transition form factors  \label{sssec:axial}}

In recent work by the Dortmund group, neutrinoproduction of the $\PPP$
$\Delta$ resonance \cite{Pas03,Lal05} was extended to cover also the
second resonance region \cite{Lal06}, which includes three isospin-1/2
resonances: the $\PP$ Roper resonance, and the two negative-parity
states $\DD$ and $\SS$.
In the following we summarize the weak transition form factors for
these resonances.
The definitions and notations for the cross sections and transition
form factors are taken from Eqs.~(IV.12)--(IV.15) and (IV.26)--(IV.28)
in Ref.~\cite{Lal06}.

\subsubsection*{\underline{$P_{33}(1232)$ resonance}}

Historically, the $P_{33}(1232)$ ($\Delta$) isobar has been studied
more extensively than any other nucleon resonance.
Electroproduction data on differential and integrated cross sections
have been used to extract the $N \to \Delta$ transition form factor,
and the resulting vector form factors, in the region $Q^2<3.5 \GeV^2$,
can be parameterized (in the notation of Ref.~\cite{Lal06}) as
\begin{equation}
C_3^{(p)}=\frac{2.13\ D_V}{1+Q^2/4 M_V^2}\ ,
\qquad
C_4^{(p)}=\frac{-1.51\ D_V}{1+Q^2/4 M_V^2}\ ,
\qquad
C_5^{(p)}=\frac{0.48\ D_V}{1+Q^2/0.776 M_V^2}\ ,
\label{eq:P1232-ff}
\end{equation}
where $D_V=1/(1+Q^2/M_V^2)^2$ is the dipole function with the vector
mass parameter $M_V=0.84\GeV$, and the superscript $(p)$ denotes a
proton target.
From isospin invariance, the electroproduction amplitudes of any
isospin-3/2 resonance, $R^{(3)}$, are equivalent for proton and
neutron targets, so that
${\cal A}(\gamma n\to R^{(3)0}) = {\cal A}(\gamma p \to R^{(3)+})$.
Since the amplitudes are linear combinations of the form factors,
the proton and neutron electromagnetic form factors are therefore
also equal, $C_i^{(n)}=C_i^{(p)}, i=3,4,5.$

The weak vector form factors $C_i^V (Q^2)$ for the amplitude
${\cal A}(W^+ n \to R^{(3)+})$ are related to the electromagnetic
form factors. For isospin-3/2 resonances, these in fact coincide,
\begin{equation}
C_i^V = C_i^{(n)} = C_i^{(p)}\ ,\quad  i=3,4,5\ .
\end{equation}
The amplitude for a proton target is related to the neutron
amplitude by Clebsch-Gordan coefficients,
${\cal A}(W^+ p \to R^{(3)++})
= \sqrt{3}\ {\cal A}(W^+ n \to R^{(3)+})$.

The form factors $C_3^{(p)}$ and $C_4^{(p)}$ from
Eq.~(\ref{eq:P1232-ff}) agree to within 5\% with those obtained
earlier under the assumption of magnetic dipole dominance.
Since they are deduced from data for $Q^2 < 3.5\GeV^2$, their
normalization and $Q^2$ dependence should be reliable in this region.

The axial form factors are obtained form PCAC,
\begin{equation}
C_5^A = \frac{1.2\ D_A}{1+Q^2/3 M_A^2}\ , \qquad
C_6^A = M^2 \frac{C_5^A}{m_\pi^2+Q^2}\ ,
\end{equation}
where $D_A=1/(1+Q^2/M_A^2)^2$ is the dipole term with the axial
mass $M_A=1.05 \GeV$.
For the other axial form factors, $C_{3,4}^A$,  we use the relations
\begin{equation}
C_4^A(Q^2) = -\frac14 C_5^A(Q^2) \quad \mbox{and} \quad C_3^A=0\ ,
\end{equation}
suggested by dispersion relations \cite{Adl68,Zuc71}.

The $P_{33}(1232)$ resonance is known to be dominant for low energy
neutrino scattering. The higher-mass resonances are very small for
$E_\nu<1.5\GeV$ and produce a noticeable peak in the invariant mass
distribution for $E_\nu>2-3\GeV$. The second peak is produced primarily by
the $D_{13}$ and $S_{11}$ resonances.

\subsubsection*{\underline{$D_{13}(1520)$ resonance}}

Among the isospin-1/2 resonances, $R^{(1)}$, the $D_{13}(1520)$
gives the dominant contribution in the second resonance region.
The proton form factors in this case differ from those of neutrons.
The vector part of the weak amplitude can be related to the
electromagnetic amplitudes by isospin symmetry,
\begin{equation}
{\cal A}^V(W^+ n \to R^{(1)+})
= {\cal A}(\gamma n \to  R^{(1)+}) - {\cal A}(\gamma p \to R^{(1)+})\ .
\end{equation}
Similarly, the weak vector form factors can be related to
electromagnetic ones via
\begin{equation}
C_i^V = C_i^{(n)} - C_i^{(p)}\ ,\quad i=3,4,5\ .
\label{eq:iV-1/2}
\end{equation}
The $Q^2$ dependence of the vector form factors (for $Q^2<3.5 \GeV^2$) was determined in
Ref.~\cite{Lal06} from precise electromagnetic data from JLab
in the second resonance region \cite{Bur02,Azn04,Azn05},
\begin{eqnarray}
D_{13}(1520): \quad
& &
C_3^{(p)} = \frac{2.95\ D_V}{1+Q^2/8.9 M_V^2}\ ,  \quad
C_4^{(p)} = \frac{-1.05\ D_V}{1+Q^2/8.9 M_V^2}\ , \quad
C_5^{(p)} = -0.48\ D_V\ ,       \nonumber\\
& &
C_3^{(n)} = \frac{-1.13\ D_V}{1+Q^2/8.9 M_V^2}\ , \quad
C_4^{(n)} = \frac{0.46\ D_V}{1+Q^2/8.9 M_V^2}\ ,  \quad
C_5^{(n)} = -0.17\ D_V\ ,       \nonumber\\
& &
\label{eq:ff-D1520}
\end{eqnarray}
for protons and neutrons, respectively.

The normalization of the axial form factors is determined by PCAC
and decay rates of the resonances.
Unfortunately, their $Q^2$ dependence cannot be determined
from the available data.
In practice, we therefore consider two cases:
(i) ``fast fall-off'', in which the $Q^2$ dependence is the same
as for the $P_{33}$ resonance,
\begin{equation}
C_5^A = \frac{-2.1\ D_A}{1+Q^2/3 M_A^2}\ ,\ \
C_6^A = M^2 \frac{C_5^A}{m_\pi^2+Q^2}\ \ \
        \mbox{(``fast fall-off'')}\ ;
\end{equation}
and (ii) ``slow fall-off'', in which the $Q^2$ dependence is flatter
and has the same form as that for the vector form factors for each
resonance,
\begin{equation}
C_5^A = \frac{-2.1\ D_A}{1+Q^2/8.9 M_A^2}\ , \ \
C_6^A = M^2 \frac{C_5^A}{m_\pi^2+Q^2}\ \ \
        \mbox{(``slow fall-off'')}\ .
\end{equation}
The other two form factors, $C_{3,4}^A$, are unknown,
and for simplicity we set them to zero, $C_3^A = C_4^A = 0$.

\subsubsection*{\underline{$P_{11}(1440)$ and $S_{11}(1535)$ resonances}}

The two lowest-lying spin-1/2 resonances, $P_{11}(1440)$ and
$S_{11}(1535)$, both have isospin $I=1/2$.
Their electromagnetic interaction depends only on two nonzero
form factors, $g_1$ and $g_2$.
For the proton these are determined for $Q^2<3.5 \GeV^2$ from electroproduction helicity
amplitudes, in analogy with the $D_{13}$ resonance,
\begin{eqnarray}
P_{11}(1440):   \quad
g_1^{(p)} &=& \frac{2.3\ D_V}{1+Q^2/4.3 M_V^2}\ ,   \nonumber\\
g_2^{(p)} &=& -0.76\ D_V
             \left[1 - 2.8 \ln\left(1+\frac{Q^2}{1\GeV^2}\right)
             \right]\ ,
\label{eq:ff-P1440}
\end{eqnarray}
and
\begin{eqnarray}
S_{11}(1535):   \quad
g_1^{(p)} &=& \frac{2.0\ D_V}{1+Q^2/1.2 M_V^2}
             \left[1+7.2\ln\left(1+ \frac{Q^2}{1\GeV^2}\right)
             \right]\ ,                     \nonumber\\
g_2^{(p)} &=& 0.84\ D_V\
              \left[1 + 0.11 \ln\left(1+\frac{Q^2}{1\GeV^2}\right)
              \right]\ .
\label{eq:ff-S1535}
\end{eqnarray}
For the neutron case, if one neglects the isoscalar contribution
to the electromagnetic current, one can use the relation
${\cal A}_{1/2}^{(n)} = -{\cal A}_{1/2}^{(p)}$.
In this case the general relation in Eq.~(\ref{eq:iV-1/2}) between
the weak and electromagnetic isovector form factors gives
$g_i^V = -2 g_i^{(p)}, i=1,2$.

The axial vector form factors of these two resonances are
constrained by PCAC,
\begin{equation}
g_3^A = g_1^A\ \frac{M(M_R\pm M)}{Q^2+m_\pi^2}\ ,
\end{equation}
with the $\pm$ corresponding to the $P_{11}$ and $S_{11}$
resonances, respectively.
At $Q^2=0$ the couplings are also determined from PCAC and the elastic
vertex of the resonance decay, which is known from experiment,
\begin{eqnarray}
P_{11}(1440): \quad
g_1^A(Q^2) &=& \frac{-0.51\ D_A}{1+Q^2/3 M_A^2}\ \ \
                \mbox{(``fast fall-off'')}\ ,           \nonumber\\
g_1^A(Q^2) &=& \frac{-0.51\ D_A}{1+Q^2/4.3 M_A^2}  \;
                \mbox{(``slow fall-off'')}\ ,
\end{eqnarray}
and
\begin{eqnarray}
S_{11}(1535): \quad
g_1^A(Q^2) &=& \frac{-0.21\ D_A}{1+Q^2/3 M_A^2}\ \ \
                \mbox{(``fast fall-off")}\ ,            \nonumber\\
g_1^A(Q^2) &=& \frac{-0.21\ D_A}{1+Q^2/1.2 M_A^2}
         \left[1+7.2\ln\left(1+\frac{Q^2}{1\GeV^2}\right)\right]\ \ \
                \mbox{(``slow fall-off")}\ .        \nonumber\\
       & &
\end{eqnarray}

\subsection{Leading twist structure functions \label{ltsf} }

The second set of inputs required for duality studies are the
inclusive structure functions $F_1 = M W_1$, $F_2 = \nu W_2$,
and $F_3 = \nu W_3$ which describe the DIS region.
Here we summarize the relevant expressions for the structure functions
in terms of leading twist parton distribution functions (PDFs).
In practice we use several PDF parametrizations, namely from the
GRV~\cite{GRV}, CTEQ~\cite{CTEQ} and MRST~\cite{MRST} groups.

For electron scattering, the $F_2$ structure function of the nucleon,
defined as the average of the proton and neutron structure functions,
is given (at leading order in $\alpha_s$ and for three flavors), by
\begin{equation}
F_2^{eN} = \frac12 \left( F_2^{ep} + F_2^{en} \right)
= \frac{5x}{18}
  \left(u + \bar{u} + d + \bar{d} + \frac{2}{5}s + \frac{2}{5}\bar{s}
  \right)\, ,
\label{eq:F2eN}
\end{equation}
where the quark distributions are defined to be those in the proton.
For neutrino scattering, the corresponding $F_2$ structure function
is given by
\begin{eqnarray}
F_2^{\nu N} &=& x (u + \bar{u} + d + \bar{d} + s + \bar{s})\ .
\label{eq:F2nuN}
\end{eqnarray}
%
%
In the moderate and large-$x$ region, where strange quarks
are suppressed, the weak and electromagnetic $F_2$ structure functions
approximately satisfy the ``5/18 rule'',
\begin{equation}
F_2^{eN}\ \approx\ \frac{5x}{18} 
\left( u+\bar{u}+d+\bar{d} \right) \
      \approx\ \frac{5}{18}\, F_2^{\nu N}\ .
\end{equation}
The experimental confirmation of the factor $5/18$ was indeed one of
the important milestones in the acceptance of the description of DIS
in terms of universal PDFs.

In the $Q^2 \to \infty$ limit, the $F_1$ structure function is related
to $F_2$ via the Callan-Gross relation, $F_2 = 2xF_1$.
Deviations from this relation arise due to perturbative $\alpha_s$
corrections, as well as from target mass effects and higher twists.
It is sometimes convenient also to define the longitudinal structure
function $F_L$,
\begin{equation}
F_L = \left( 1 + \frac{4 M^2 x^2}{Q^2} \right) F_2 - 2x F_1\ .
\label{eq:F_L}
\end{equation}
For large $Q^2$ the target mass term proportional to $M^2/Q^2$ can be
omitted; however, at $Q^2 \sim$~few GeV$^2$ it can make an important
contribution, especially at large $x$.
Because the extraction of $F_L$ requires longitudinal--transverse
separation of cross section data, which is challenging
experimentally, in practice the $F_L$ structure function is not
very well determined. For $F_L$ we use the parametrization of the
MRST group \cite{MRST}. To estimate the uncertainty in its determination, we consider
two different scenarios for $2xF_1$, namely
(i) Callan-Gross relation, $2xF_1=F_2$, and
(ii) the exact expression for $2xF_1$ from Eq.~(\ref{eq:F_L}).

Finally, the charge-conjugation odd $F_3$ structure function for
neutrino-nucleon scattering is given by
\begin{eqnarray}
xF_3^{\nu p} = 2x (d - \bar{u} + s), \quad  & xF_3^{\nu n} = 2x (u -
\bar{d} + s) \ .
\end{eqnarray}
%
%
%
%
%
%
If one neglects the contribution of the strange quarks, $s\approx 0$,
which is the case we consider here, the isoscalar $F_3^{\nu N}$ structure
function is given simply by the valence $u_v$ and $d_v$ distributions:
\begin{eqnarray}
xF_3^{\nu N} & \approx & x((u- \bar{u})+(d-\bar{d}))  =  x (u_v + d_v)\ .
\end{eqnarray}
In the large-$x$ region, where contribution of all sea quarks is very
small,  the $F_2$ structure function will also be proportional to $xF_3$,
\begin{eqnarray}
xF_3^{\nu N} & \approx & F_2^{\nu N}  \approx  \frac{18}{5} F_2^{eN}\ .
\end{eqnarray}

In the next section we will consider duality both for the total structure
function, and for the valence-only structure function. In the context of
``two--component duality'' \cite{Har69}, the resonance contributions are
taken to be dual to valence quarks, while the nonresonant background is
dual to the sea. In the resonance region, and especially at low $Q^2$, it
may be reasonable that a resonance-based model of structure functions
would generate a valence-like scaling function. Indeed, there were strong
suggestions of such resonance-valence duality in the recent proton $F_2$
data from JLab \cite{Nic00}. In the present study we test the
``two-component'' duality hypothesis by comparing the calculated resonance
structure functions with both the total and valence-only structure
functions.

In the next section we will use the above expressions to quantify
the degree to which the averaged resonance structure function duals
the leading twist structure functions for neutrino scattering,
and compare this with duality for the electron case.

\section{Duality in electron \& neutrino structure functions
\label{LocalDuality}}

\subsection{Electron scattering}

Before proceeding with the discussion of duality in neutrino scattering,
we first consider duality for the better known case of electron
scattering. Recent high-precision experiments at Jefferson Lab and
elsewhere
\cite{Nic00,Chr02,Arr03,Liang04,Wesselmann:2006mw,Bur01,JLa01,Air03} have
allowed accurate tests to be performed of Bloom-Gilman duality in electron
scattering. For the proton $F_2$ structure function, Niculescu {\em et
al.} \cite{Nic00} found that the structure function in the resonance
region, averaged over several intervals of $x$ corresponding to the
prominent resonance regions, reproduces well the scaling structure
function down to relatively low values of $Q^2$. Our aim here will not
necessarily be to reproduce accurately the data with our resonance model
\cite{Pas03,Lal05,Lal06}, but rather using phenomenological information on
transition form factors to compare the workings of duality for neutrinos
and electrons.

\begin{figure}[h!bt]
\includegraphics[angle=-90,width=0.49\textwidth]{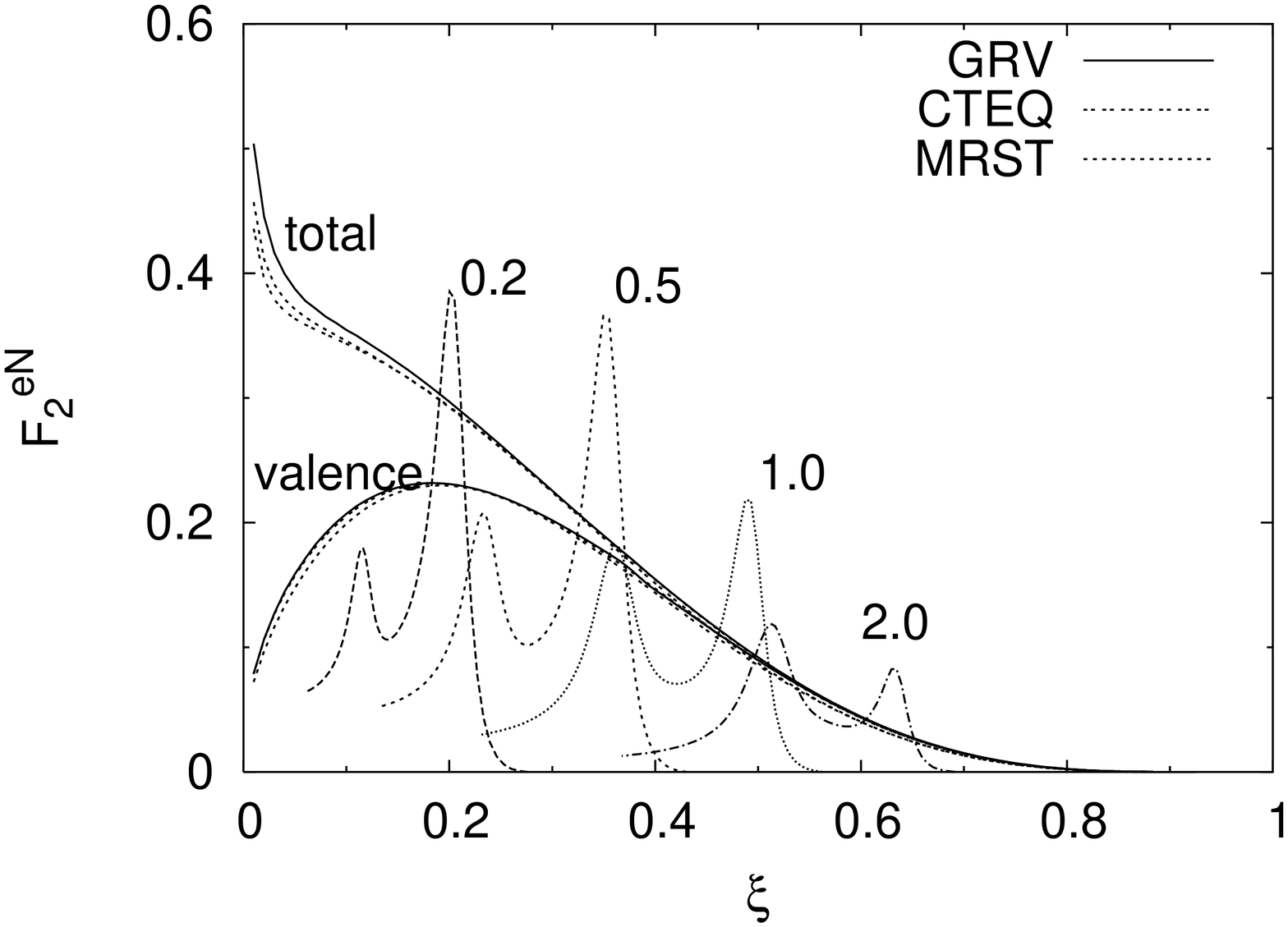}
\includegraphics[angle=-90,width=0.49\textwidth]{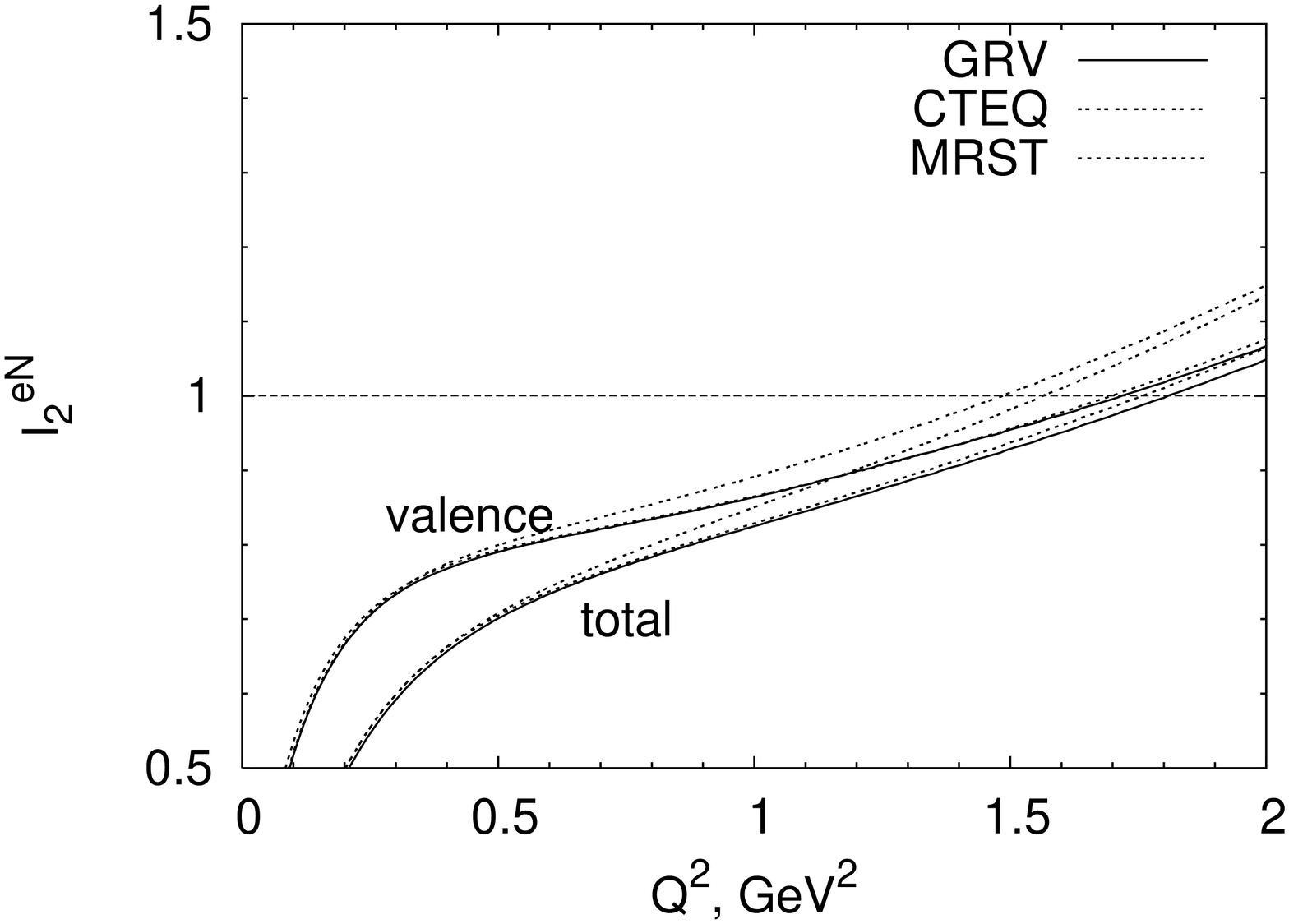}
\caption{
    Duality for the isoscalar nucleon $F_2^{eN}$ structure function.
{\em (Left)}
    $F_2^{eN}$ as a function of $\xi$, for $Q^2 = 0.2, 0.5, 1$
    and $2\GeV^2$ (indicated on the spectra), compared with several leading twist
    parameterizations \cite{GRV,CTEQ,MRST} (valence and total)
    at $Q^2=10\GeV^2$.
{\em (Right)}
    Ratio $I_2^{eN}$ of the integrated $F_2^{eN}$ in the resonance
    region to the leading twist functions (valence and total).}
\label{fig:F2-eN-Na}
\end{figure}

The isoscalar nucleon structure function
$F_2^{eN} = (F_2^{ep} + F_2^{en})/2$,
calculated as a sum of electroproduced resonances, is displayed
in Fig.~\ref{fig:F2-eN-Na} as a function of the Nachtmann variable
$\xi = 2x/\sqrt{1 + 4 M^2 x^2/Q^2}$ for several values of $Q^2$
from 0.2 to 2~GeV$^2$.
The use of the Nachtmann variable takes into account kinematical target
mass corrections, which can be important at large $x$ and low $Q^2$.
The prominent peaks correspond to the $P_{33}(1232)$ ($\Delta$)
resonance at the largest $\xi$ values in each spectrum.
The next peaks, at smaller $\xi$, correspond to the second resonance
region, where the $S_{11}(1535)$ and $D_{13}(1520)$ resonances
dominate, and the $P_{11}(1440)$ resonance gives a small contribution.
With increasing $Q^2$, the resonance peaks decrease in height and
move to larger $\xi$.

To examine the extent to which ``local duality'' works, we compare the
$\xi$ and $Q^2$ dependence of the individual resonances with the $\xi$
dependence of the leading twist $F_2^{eN}$ structure function, for both
the total and the valence--only cases. For the latter, we use leading
twist PDFs at $Q^2=10\GeV^2$ from the GRV \cite{GRV}, CTEQ \cite{CTEQ} and
MRST \cite{MRST} groups. On average the resonances appear to oscillate
around and slide down the leading twist function, reminiscent of the
general features of the data as a function of $\xi$ and $Q^2$ --- see
Refs.~\cite{Nic00,Mel05}.
For the calculated structure function, we consider the four resonances
mentioned above, and integrate over the region
\begin{equation}
1.1 \leq W \leq 1.6 \GeV \, ,
\label{eq:Wrange}
\end{equation}
where the upper bound covers the range of the resonances taken into
account in this analysis.

The degree to which local duality is valid can be quantified by
considering the ratio of integrals of the resonance (res) and
leading twist (LT) structure functions,
\begin{equation}
I_i(Q^2) =
\frac{ \int_{\xi_{\rm min}}^{\xi_{\rm max}} d\xi\
    {\cal F}_i^{(\rm res)}(\xi,Q^2) }
     { \int_{\xi_{\rm min}}^{\xi_{\rm max}} d\xi\
    {\cal F}_i^{(\rm LT)}(\xi,Q^2) }\ ,
\label{eq:Int}
\end{equation}
where ${\cal F}_i$ denotes $F_2$, $2xF_1$ or $xF_3$,
and the integration limits correspond to
$\xi_{\rm min} = \xi(W=1.6~{\rm GeV},\, Q^2)$ and
$\xi_{\rm max} = \xi(W=1.1~{\rm GeV},\, Q^2$).
The closer this ratio is to unity, the better the agreement
with duality will be.
Defining the ratio $I_i(Q^2)$ in terms of integrals over the Nachtmann
scaling variable $\xi$ instead of Bjorken $x$ implicitly includes
target mass corrections in the structure functions
\cite{Geo76,Kre03,Ste06}, which are important at large $x$ and small
$Q^2$.
This is especially so for the $F_L$ structure function, which is
intrinsically small.
An alternative approach would be to express the target mass corrected
structure functions in terms of $x$ and $Q^2$ \cite{Geo76} and perform
the integrations over $x$ \cite{Mel05}.
For a first investigation of duality, and since we are mostly concerned
about the relative differences between duality in neutrino and electron
scattering, the integrals over $\xi$ in Eq.~(\ref{eq:Int}) will provide
a sufficient test of integrated duality.

The ratio $I_2^{eN}$ for electron scattering is shown in
Fig.~\ref{fig:F2-eN-Na} (right panel).
The results are similar to those of the empirical analysis of JLab
proton data \cite{Nic00}.
The integrated resonance contribution is smaller than the leading
twist at low $Q^2$, but increases with increasing $Q^2$.
For $Q^2 \gtrsim 1\GeV^2$, the ratio $I_2^{eN}$ is within $\sim 20\%$
of unity when using the total DIS structure function.
On the other hand, for the valence-only structure function the ratio
is within $\sim 20\%$ of unity over a larger range,
$Q^2 \gtrsim 0.5\GeV^2$.
The better agreement of the resonance curve with the valence-only
leading twist curve supports the notion of two--component duality \cite{Har69},
as observed in the JLab $F_2^{ep}$ data \cite{Nic00}.
In more refined treatments one would also take into account the $Q^2$
evolution of the leading twist structure function.
This will modify the quantitative behavior of the ratio with respect
to $Q^2$, but not its essential features.

The fact that $I_2^{eN}<1$ in our model can be understood from the
fact that only the first four resonances are included in the structure
function.
Since $F_2$ is positive, the contribution from higher resonances as
well as the nonresonant background will increase the numerator in
the ratio $I_2^{eN}$ and thus improve the accuracy of duality.
The behavior of the ratio $I_2^{eN}$ at large $Q^2$ is less well
constrained due to the current poor knowledge of the leading twist
structure function at high $x$ and of the transition form factors
at large $Q^2$.

\begin{figure}[h!bt]
\includegraphics[angle=-90,width=0.49\textwidth]{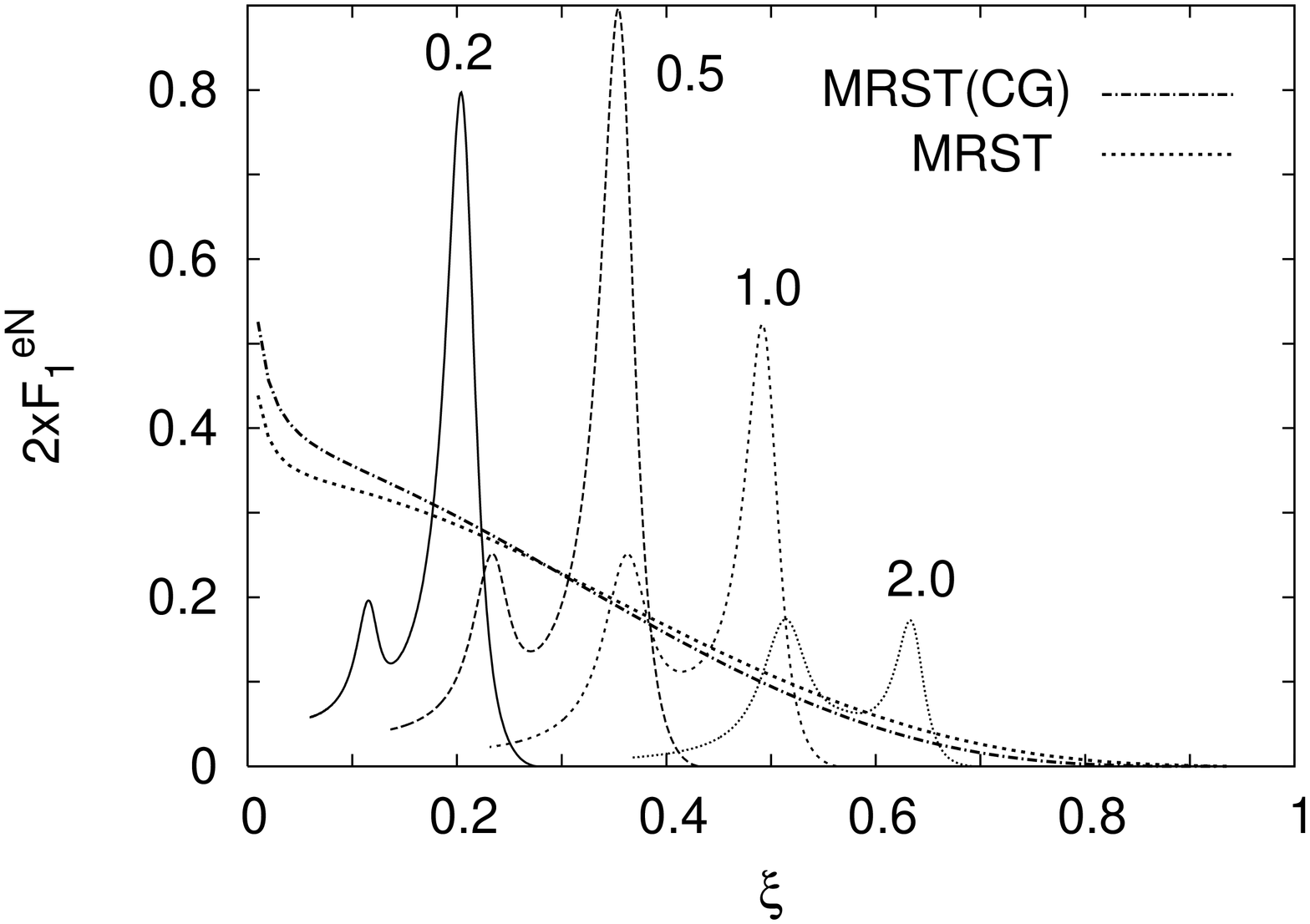}
\includegraphics[angle=-90,width=0.49\textwidth]{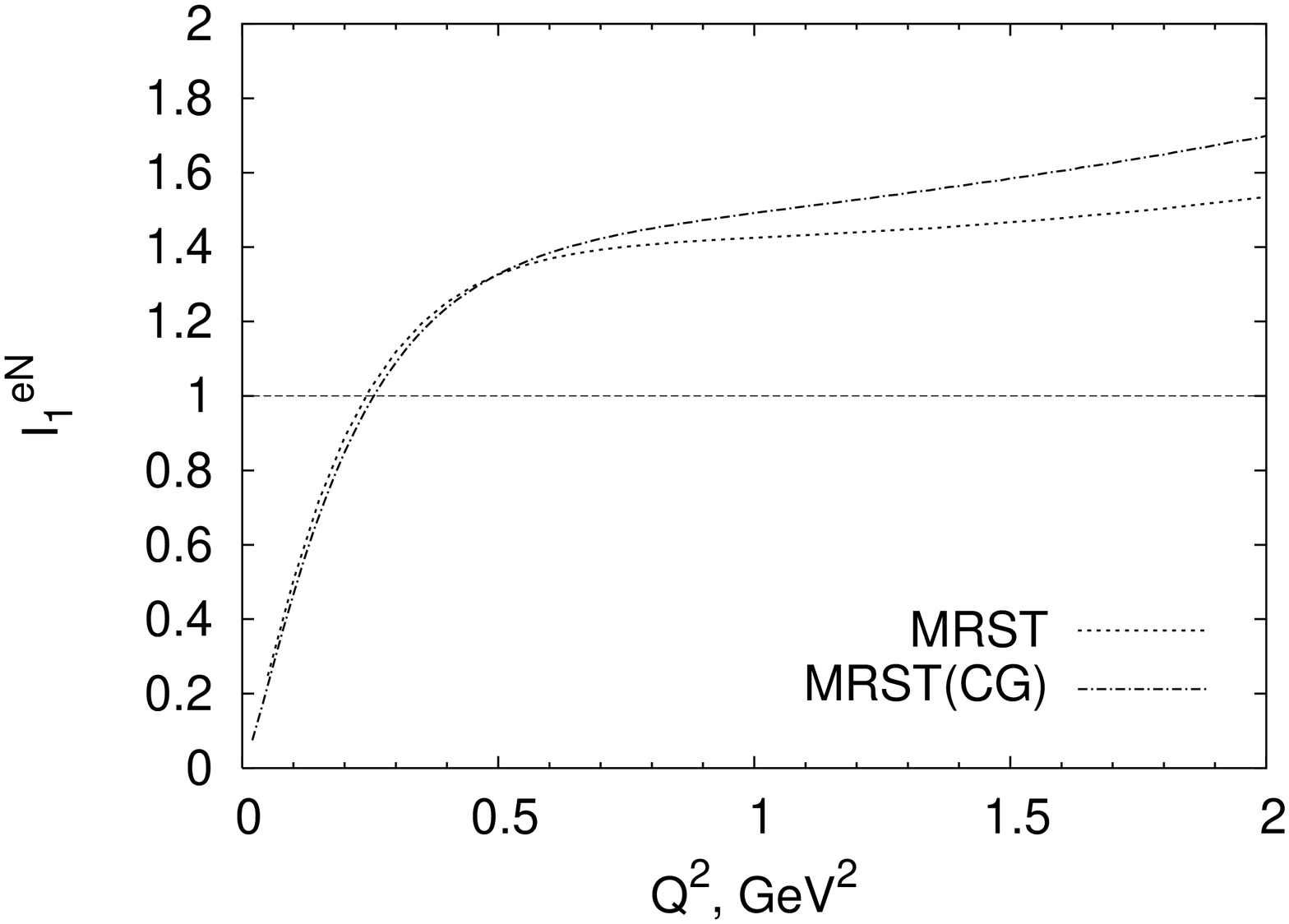}
\caption{
    Duality for the isoscalar nucleon $2x F_1^{eN}$ structure
    function.
{\em (Left)}
    $2x F_1^{eN}$ as a function of $\xi$, for $Q^2 = 0.2, 0.5, 1$
    and $2\GeV^2$ (indicated on the spectra), compared with the MRST
    parameterization \cite{MRST} at $Q^2=10\GeV^2$, using Eq.(\ref{eq:F_L})
    (dotted) and the Callan--Gross (CG) relation, $F_2=2xF_1$ (dot-dashed).
{\em (Right)}
    Ratio $I_1^{eN}$ of the integrated $2x F_1^{eN}$ in the
    resonance region to the leading twist function \cite{MRST}
    (see text).}
\label{fig:F12x-eN-Na}
\end{figure}

Recently, new high-precision data from Jefferson Lab have allowed
longitudinal-transverse separations to be performed, which have enabled
the proton $2x F_1$ structure function to be accurately determined at
large $x$ \cite{Liang04}. This has made it possible for the first time to
perform quantitative tests of duality for the $F_1$ (or $F_L$) structure
function. In Fig.~\ref{fig:F12x-eN-Na} (left panel) we plot the isoscalar
nucleon structure function $2x F_1^{eN}$, calculated for the above
mentioned four resonances, and compare with the leading twist
parametrization from Ref.~\cite{MRST} at $Q^2=10$~GeV$^2$.
The two leading twist curves correspond to the two scenarios for $2xF_1$
discussed in Sec.~\ref{ltsf}, namely, using the Callan-Gross relation,
$F_2 = 2x F_1$, and using the exact expression in Eq.~(\ref{eq:F_L}).
The difference between the two curves is relatively small at
$Q^2=10$~GeV$^2$, so that one can use either in the comparison
with the resonance structure function.

With increasing $Q^2$, the resonance $2x F_1^{eN}$ structure function
is seen to slide along the leading twist curve, just as in the case of $F_2^{eN}$,
but on average sits slightly higher than the leading twist curve.
This can be quantified by considering the ratio $I_1^{eN}$, defined
in Eq.~(\ref{eq:Int}), which we plot in Fig.~\ref{fig:F12x-eN-Na}
(right panel).
For most of the range of $Q^2 \gtrsim 0.5$~GeV$^2$ the ratio is some
$30-50\%$ above unity, which may indicate the need for additional
terms in the resonance sum.
On the other hand, it is known that target mass corrections have a
relatively larger effect on $2x F_1$ than on $F_2$ \cite{Mel05,Geo76},
and would tend to increase the leading twist functions, especially at
large $x$ (or $\xi$), and hence reduce the ratio $I_1^{eN}$.

\subsection{Neutrino scattering}

In the previous section we have demonstrated that the resonance model used here \cite{Lal06}
reproduces the qualitative features of duality observed in electron
scattering, and have established the accuracy with which this duality
holds in the model.
Here we turn to the main aim of our paper, which is to compare and
contrast the workings of duality in $eN$ scattering and in $\nu N$
scattering.
To make the comparison as rigorous as possible, we calculate the
neutrino structure functions using the same four resonances as for
the electron case, Figs.~\ref{fig:F2-eN-Na}--\ref{fig:F12x-eN-Na}.

Neutrino interactions have particular features which distinguish them
from electromagnetic probes.
For the charge current reaction $\nu_\mu \, p \to \mu^- \, \Delta^{++}$,
for example, only isospin-3/2 resonances are excited, and in particular
the $P_{33}(1232)$ resonance.
Because of isospin symmetry constraints, the neutrino--proton structure
functions ($F_2^{\nu p}$, $2xF_1^{\nu p}$ and $xF_3^{\nu p}$) for these
resonances are three times larger than the neutrino--neutron structure
functions.
In this case the resonance structure functions are significantly larger
than the leading twist functions,
$F_i^{\nu p {\rm (res)}} > F_i^{\nu p {\rm (LT)}}$,
and quark-hadron duality is clearly violated for a proton target.

In neutrino--neutron scattering, in addition to isospin-3/2 resonances,
all the isospin-1/2 resonances can also be excited.
However, the total contribution of the three isospin-1/2 resonances
considered here is smaller than that from the leading $P_{33}(1232)$
resonance.
The leading twist curve for the $\nu n$ structure functions lies above
the resonance structure functions,
$F_i^{\nu n {\rm (res)}} < F_i^{\nu n {\rm (LT)}}$,
so that quark-hadron duality does not hold for this case either.

The general feature of the resonance curves is that at the onset of the
resonance region, $W \lesssim 1.6\GeV$, the neutrino--proton structure
functions are larger than the corresponding neutrino--neutron ones.
In the deep inelastic region, on the other hand, the structure
functions are larger for neutrino--neutron scattering.
It has been argued \cite{Clo01} for the case of electron scattering
that for duality to appear one must sum over a complete set of even
and odd parity resonances.
In neutrino scattering, due to isospin symmetry constraints, duality
will not hold locally for protons and neutrons separately, even if
several resonances with both even and odd parities are taken into
account \cite{Clo03}.
In this case one can consider duality for the average of proton
and neutron structure functions, which is the approach we take in
this work.
We will demonstrate below that in this case duality holds with even
greater accuracy than for electron scattering.

This discussion raises the question of how the transition occurs from
the resonance to DIS regions in the case of neutrino scattering.
We can speculate about the possible mechanisms of how this takes place.
Starting from low $W$, the first resonance is the $P_{33}(1232)$, whose
contribution to the $\nu p$ structure function is three times larger
than that to the $\nu n$, as mentioned above.
To compensate its influence, this resonance must be followed by several
isospin-1/2 resonances, which can only contribute to neutrino--neutron
structure functions.
This is what indeed happens --- the $\PP$, $\DD$ and $\SS$ resonances
are the next ones in the mass spectrum.
In fact, the results of Ref.~\cite{Lal06} show that the $P_{33}(1232)$
form factors fall faster that a dipole, while those for the
$D_{13}(1520)$ and $S_{11}(1535)$ resonances fall slower than a dipole,
so that at $Q^2 \approx 2\GeV^2$ the two peaks are comparable.
From our calculations we also know that with only these resonances the
$\nu n$ cross section (and structure functions) are still smaller than
those of $\nu p$.
Additional resonances with higher masses may also follow this trend
and further enhance the $\nu n$ structure functions to overcome those
for $\nu p$.

At higher masses the isospin-3/2 resonances $P_{33}(1600)$ and
$S_{31}(1620)$ appear, which also give three times larger
contributions for $\nu p$ scattering than for $\nu n$.
They are again followed by the three isospin-1/2 resonances,
$S_{11}(1650)$, $D_{15}(1675)$ and $F_{15}(1680)$, two of which have
spin 5/2.
Their contributions can be large due to the summation over six final
spin states, which further increases the neutrino--neutron structure
functions.
One could suppose that in this region the neutrino--neutron
contribution would exceed the neutrino--proton.
Furthermore, we have again one isospin-3/2 resonance, the
$D_{33}(1700)$, and three isospin-1/2 resonances --- $D_{13}(1700)$,
$P_{11}(1710)$ and $P_{13}(1720)$ --- to compensate its influence.
Above $W = 1750 \MeV$, and up to $2220 \MeV$, the isospin-3/2
resonances prevail, with 11 known states, and only 9 with
isospin 1/2.

A more detailed investigation of the interplay between the resonances
with different spins would be possible after the form factors are
determined for at least some of these higher-lying resonances.
At present, however, we consider only the first four resonances,
for which the $\nu p$ cross section is always larger than the leading
twist contribution, and the $\nu n$ cross section is always smaller.
This is one additional reason to compare only the average of the
$\nu p$ and $\nu n$ structure functions.

\begin{figure}[h!bt]
\includegraphics[angle=-90,width=0.49\textwidth]{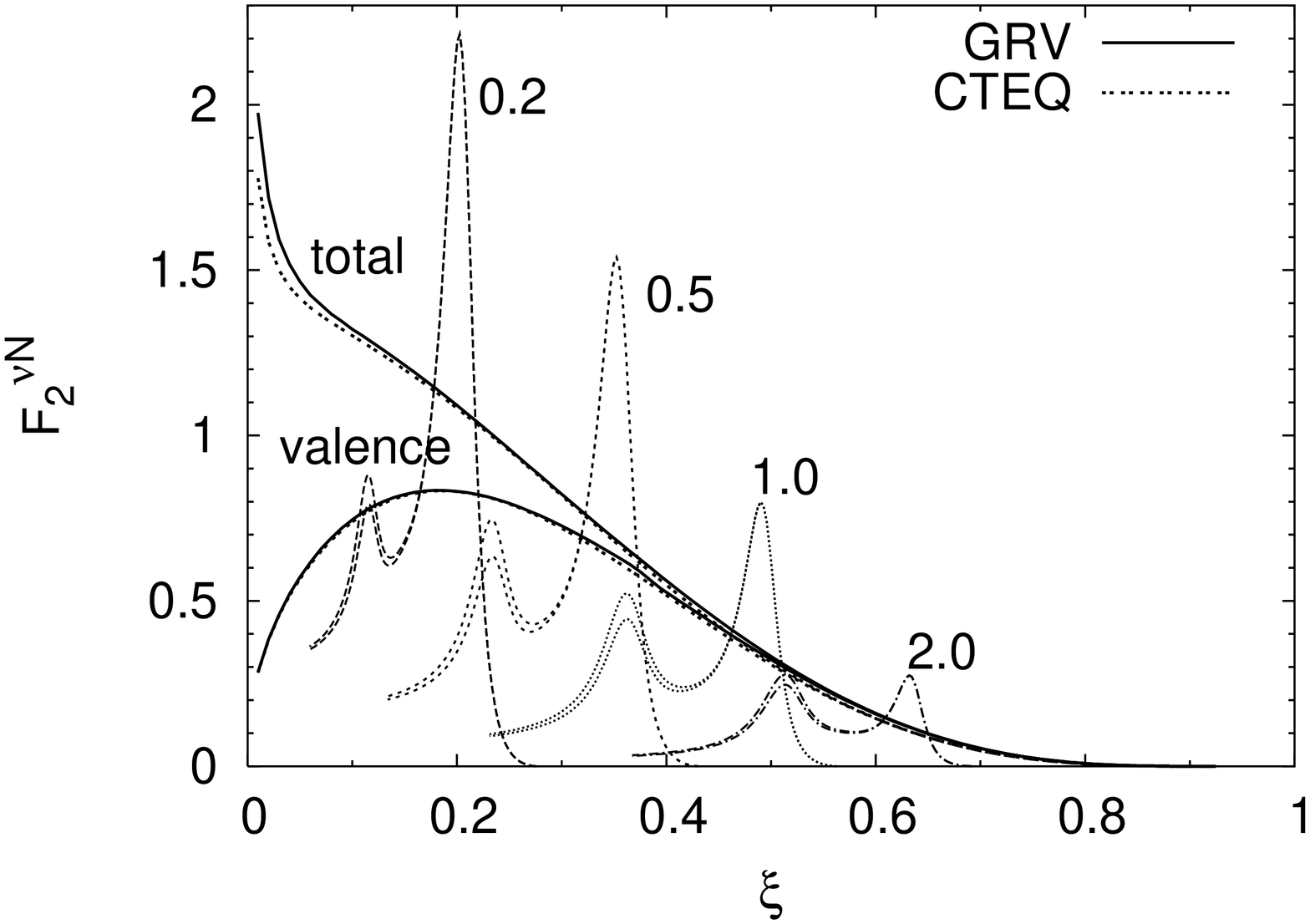}
\includegraphics[angle=-90,width=0.49\textwidth]{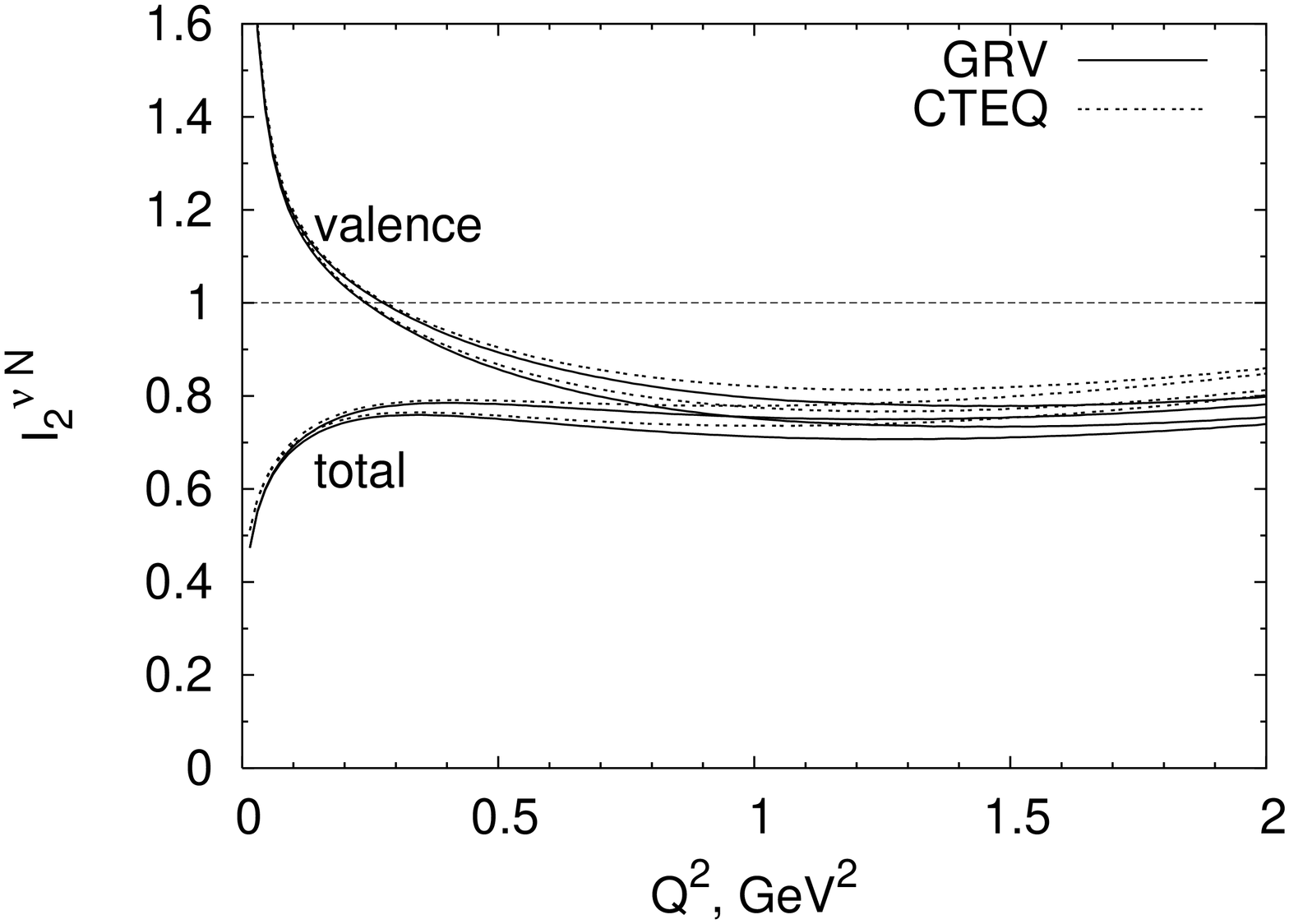}
\caption{
    Duality for the neutrino--nucleon $F_2^{\nu N}$ structure
    function.
{\em (Left)}
    $F_2^{\nu N}$ in the resonance region at several $Q^2$ values
    (indicated on the spectra), compared with leading
    twist parameterizations \cite{GRV,CTEQ}
    (valence and total) at $Q^2=10\GeV^2$.
{\em (Right)}
    Ratio $I_2^{\nu N}$ of the integrated $F_2^{\nu N}$ in the
    resonance region to the leading twist functions \cite{GRV,CTEQ} (valence and
    total). The upper (lower) resonance curves and the upper (lower) integrated
    ratios correspond to the "slow" ("fast") fall-off of the axial form factors.}
\label{fig:F2-nuN-Na}
\end{figure}

The neutrino--nucleon $F_2^{\nu N}$ structure function is displayed
in Fig.~\ref{fig:F2-nuN-Na} (left panel) as a function of $\xi$ for
several values of $Q^2$.
Here the $P_{33}(1232)$ resonance is seen as the largest peak at each
$Q^2$.
The next peak at lower $\xi$ (larger $W$) is dominated by the $\DD$
and $\SS$ resonances.
The contribution from the latter becomes more significant with
increasing $Q^2$ since its form factors fall off more slowly than
the dipole.
The contribution of the $\PP$ resonance is too small to be seen as a
separate peak.
The two sets of resonance curves correspond to the ``fast fall-off''
(lower curves) and ``slow  fall-off'' (upper curves) scenarios for the
axial form factors discussed in Sec.~\ref{sssec:axial}.
The smooth curves are obtained from Eq.~(\ref{eq:F2nuN}) using the
GRV \cite{GRV} and CTEQ \cite{CTEQ} leading twist parton distributions
at $Q^2=10\GeV^2$, as in Fig.~\ref{fig:F2-eN-Na}.
Just as in the case of electron--nucleon scattering, with increasing
$Q^2$ the resonances slide along the leading twist curve, which is
required by duality.
As in Fig.~1, we show both the total structure function and the
valence-only contribution.

In Fig.~\ref{fig:F2-nuN-Na} (right panel) we show the ratio of the
integrals of the neutrino resonance and leading twist structure
functions, defined in Eq.~(\ref{eq:Int}).
The ratio is within $\sim$~20--25\% of unity for $Q^2 \gtrsim 0.3\GeV^2$
and, unlike the corresponding electron--nucleon ratio $I_2^{eN}$, does
not grow appreciably with $Q^2$.
Again, the two sets of resonance curves correspond to the
``fast fall-off'' (lower) and ``slow  fall-off'' (upper) scenarios
for the axial form factors.
The difference between the curves reflects the uncertainty in the
calculation of $I_2^{\nu N}$.
As expected, this ratio is close to 1 for the ``valence-only'' function
at low $Q^2$, which favors the hypothesis of two--component duality \cite{Har69}.
A comparison for $Q^2 \lesssim 0.5\GeV^2$ may be questionable, however,
since there the perturbative QCD expansion is unlikely to be valid.
For large $Q^2$ the ratio is sensitive to the parametrization used
for the leading twist curve, and the difference between the two
parametrizations is smaller than the difference between the valence-only
and total functions.

\begin{figure}[h!bt]
\includegraphics[angle=-90,width=0.49\textwidth]{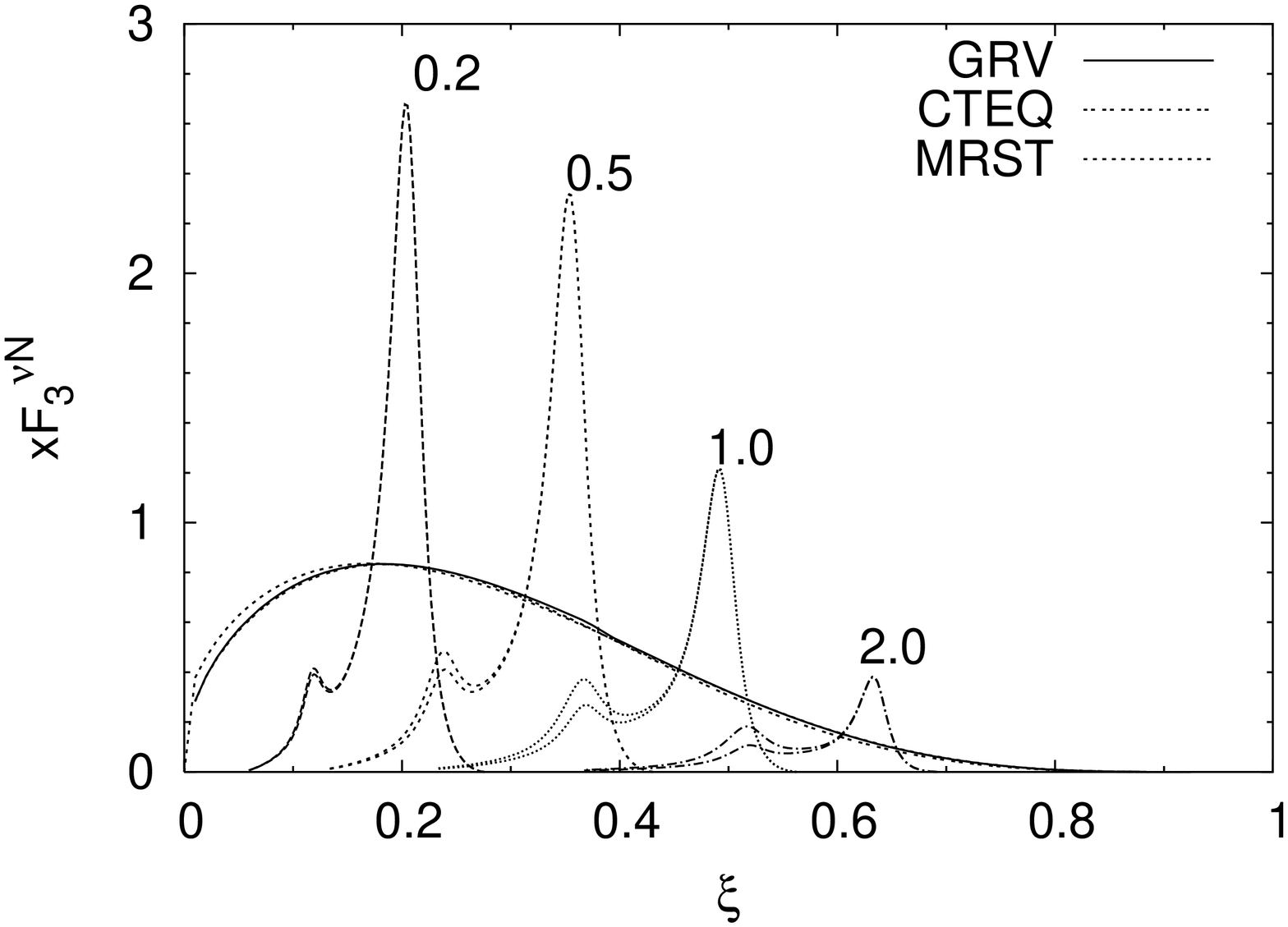}
\includegraphics[angle=-90,width=0.49\textwidth]{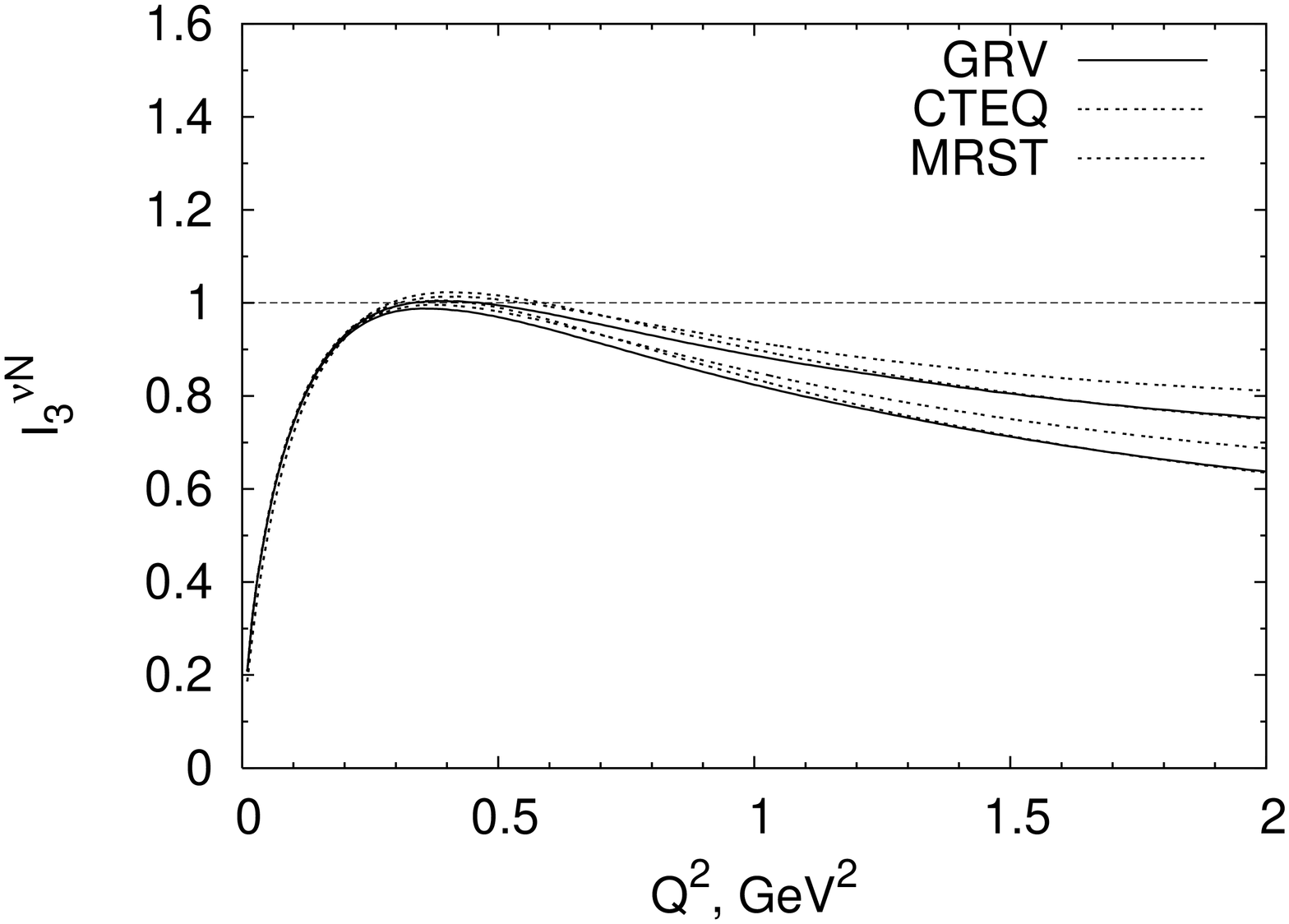}
\caption{
    Duality for the neutrino--nucleon $xF_3^{\nu N}$ structure
    function.
{\em (Left)}
    $xF_3^{\nu N}$ in the resonance region for several $Q^2$ values,
    (indicated on the spectra), compared with several
    leading twist parameterizations \cite{GRV,CTEQ,MRST} at
    $Q^2=10\GeV^2$.
{\em (Right)}
    Ratio $I_3^{\nu N}$ of the integrated $xF_3^{\nu N}$
    in the resonance region to the leading twist functions.
The upper (lower) resonance curves and the upper (lower) integrated ratios
correspond to the ``slow'' (``fast'') fall-off of the axial form
factors.}
\label{fig:xF3}
\end{figure}

New features appear when considering the $C$-odd structure function
$F_3^{\nu N}$.
As discussed above for the case of $F_2^{\nu N}$, for the resonances
considered here the proton $F_3^{\nu p}$ structure function is larger
than the neutron $F_3^{\nu n}$, whereas for deep inelastic scattering
the $\nu n$ is larger.
In our analysis we compare the isoscalar nucleon data, which are shown
in Fig.~\ref{fig:xF3} (left panel).
As before, the lower and upper curves in the second resonance region
correspond to the ``slow'' and ``fast'' fall-offs of the axial form
factors, respectively.

To quantify the degree to which the resonance and deep inelastic
structure functions are dual, we calculate the ratio of integrals
for the $xF_3^{\nu N}$ structure function as in Eq.~(\ref{eq:Int}).
This ratio, shown in Fig.~\ref{fig:xF3} (right panel), appears to fall off more
rapidly with $Q^2$ than for the $F_2^{\nu N}$ ratio, and reaches
$\sim 0.7$ at $Q^2=2\GeV^2$.
The $F_3^{\nu N}$ structure function is in general more sensitive
to the choice of axial form factors, and our results are consistent
with the uncertainty in the axial form factors, which is estimated
to be $\sim 30\%$ at $Q^2=2\GeV^2$.

\begin{figure}[h!bt]
\includegraphics[angle=-90,width=0.49\textwidth]{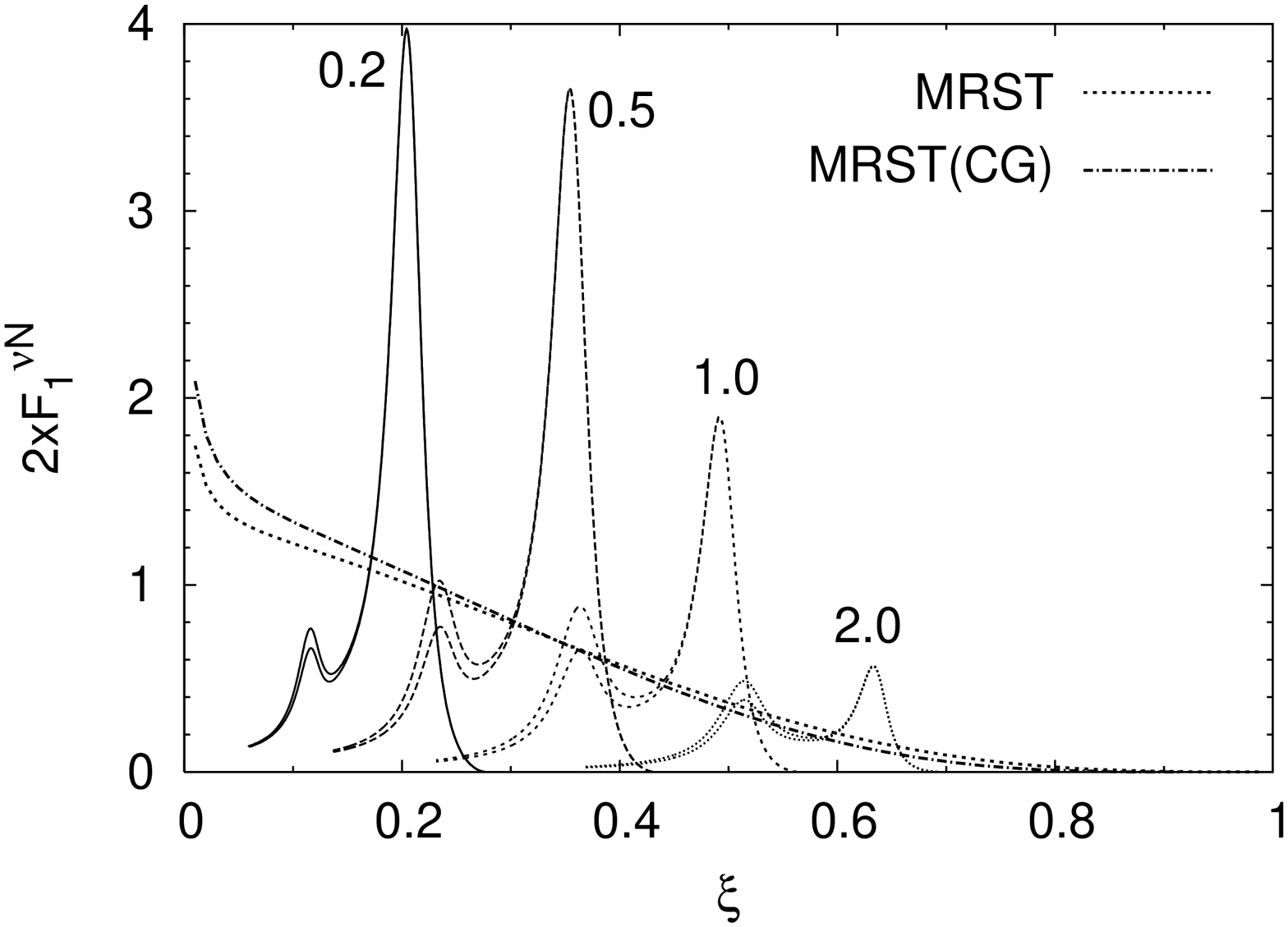}
\includegraphics[angle=-90,width=0.49\textwidth]{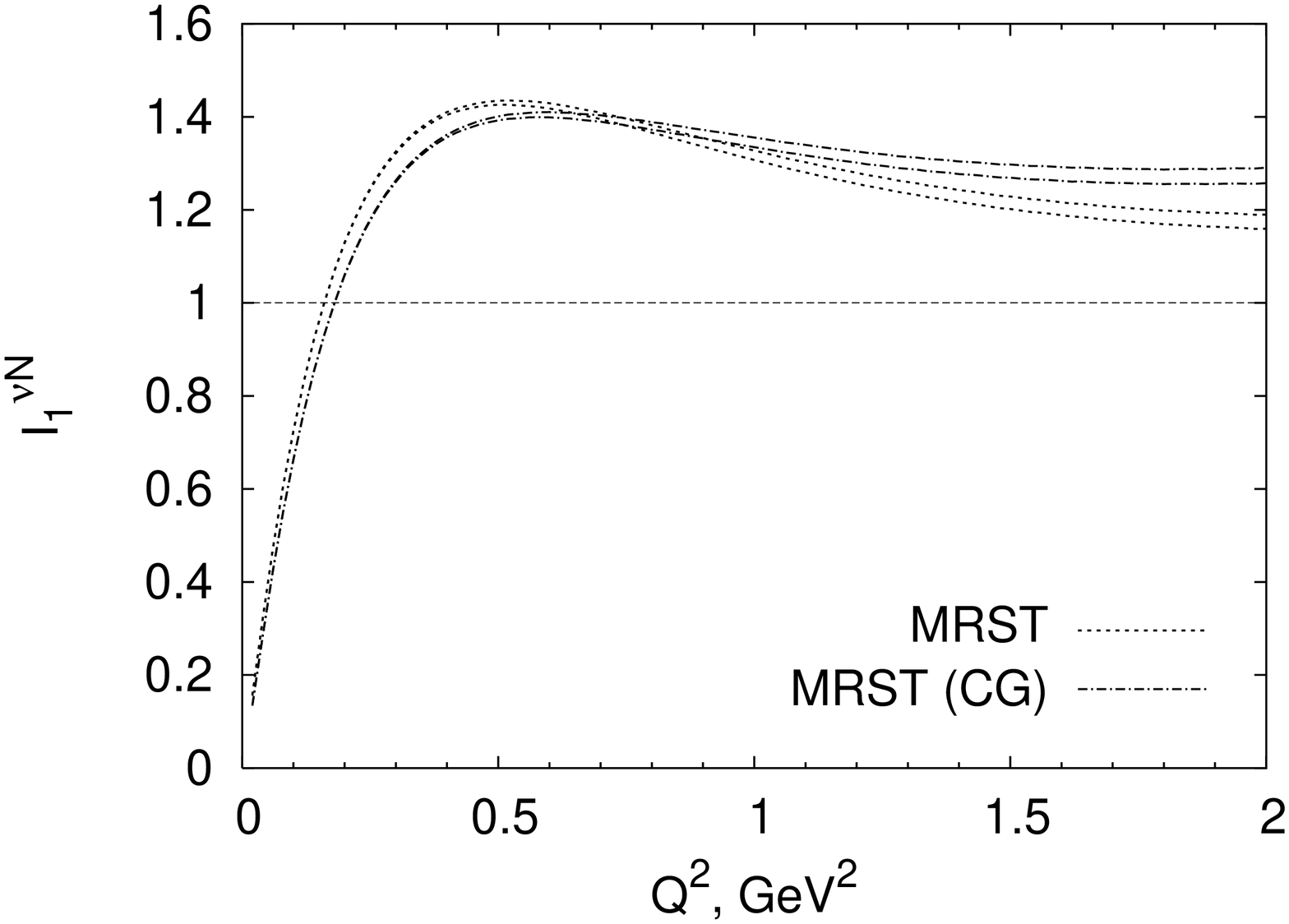}
\caption{
    Duality in the neutrino--nucleon $2x F_1^{\nu N}$ structure
    function.
{\em (Left)}
    $2x F_1^{\nu N}$ in the resonance region at several $Q^2$ values
    (indicated on the spectra), compared with
    the MRST parameterization \cite{MRST} at $Q^2=10\GeV^2$ using the exact
    expression in Eq.(\ref{eq:F_L}) (dotted) and Callan-Gross relation
    (dot-dashed).
{\em (Right)}
    Ratio $I_1^{\nu N}$ of the integrated $2x F_1^{\nu N}$ in the
    resonance region to the leading twist function \cite{MRST}.
The upper (lower) resonance curves and the upper (lower) integrated ratios
correspond to the ``slow'' (``fast'')  fall-off of the axial form
factors.}
\label{fig:F12x-nuN-Na}
\end{figure}

Finally, in Fig.~\ref{fig:F12x-nuN-Na} (left panel) we show the neutrino
structure function $2x F_1^{\nu N}$ as a function of $\xi$ for several
$Q^2$ values. The resonance structure function are calculated for "slow
fall-off" and "fast fall-off" axial form factors. The leading twist functions
correspond to the MRST parametrization \cite{MRST} using the Callan--Gross
relation and the exact expression in Eq.(\ref{eq:F_L}).
As in the electron scattering case, the resonance contributions appear
to lie above the leading twist curve for most of the range of $\xi$.
The ratio $I_1^{\nu N}$, shown in Fig.~\ref{fig:F12x-nuN-Na} (right
panel), is about 20\% above 1 for $Q^2 > 1$~GeV$^2$, which again may
be an indication that target mass effects need to be removed from the
leading twist structure function before comparing with the resonance
contributions.

\subsection{Adler sum rule \label{SumRules}}

One of the most fundamental results in neutrino scattering is the
relation between the difference of the $\nu n$ and $\nu p$
structure functions for quasi-elastic (QE) scattering and for the
rest of the higher mass states \cite{Adl65,Wei66,Adl66},
\begin{equation}
  \left[ g_{1V}^{(QE)}(Q^2) \right]^2
+ \left[ g_{1A}^{(QE)}(Q^2) \right]^2
+ \left[ g_{2V}^{(QE)}(Q^2) \right]^2 \frac{Q^2}{4 M^2}
+ \int d\nu
  \left[ W_2^{\nu n}(Q^2,\nu) - W_2^{\nu p}(Q^2,\nu) \right]
= 2\ .
\label{eq:AdlerSR}
\end{equation}
Because it measures the isospin of the target, this relation must hold for all
values of $Q^2$.

In the $Q^2 \to 0$ limit, Eq.~(\ref{eq:AdlerSR}) is reduced to the
Adler-Weisberger relation \cite{Adl65,Wei66}, which has been verified
experimentally to good accuracy.
For $Q^2\ne 0$, it is known as the Adler sum rule \cite{Adl66}, which
has also been tested with data for neutrino deep inelastic
scattering, and found to hold to $\approx 20\%$ accuracy \cite{WA25}.
At large $Q^2$ it has a simple interpretation in the parton model,
in terms of integrals of valence quark distributions.
Using the model \cite{Lal06} for the resonance form factors, we can study how the
Adler sum rule is satisfied as a function of $Q^2$.

\begin{figure}[h!tb]
\begin{center}
\includegraphics[angle=-90,width=0.80\textwidth]{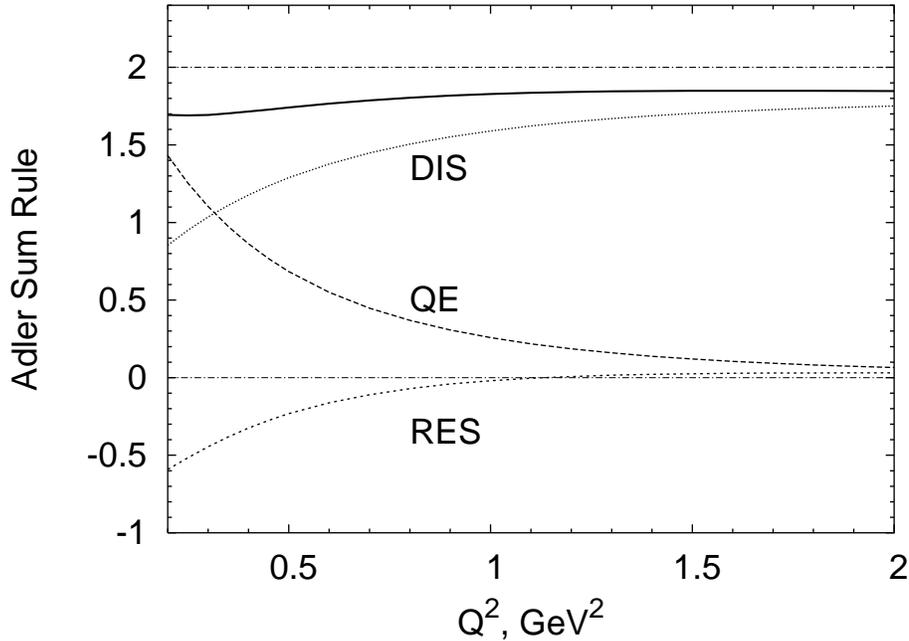}
\end{center}
\caption{Decomposition of the Adler sum, as a function of $Q^2$,
    into its QE (dashed), resonance (short dashed) and deep
    inelastic (dotted) contributions, as well as the total
    (solid).}
\label{fig:AdlerSR}
\end{figure}

For the QE form factors we use the following simple parametrization:
\begin{equation}
g_{1V}^{(QE)}=\frac1{D_V}, \quad g_{2V}^{(QE)}=\frac{3.7}{D_V}, \quad
g_{1A}^{(QE)}=\frac{1.23}{D_A}\ .
\end{equation}
The $W_2$ structure functions in Eq.~(\ref{eq:AdlerSR}) include
contributions from resonance production and from the deep inelastic
region. The resonance contribution is calculated for the first four
resonances, as discussed earlier.
The integration is performed in the range of
$\nu_{\rm min} < \nu < \nu_{\rm max}$ corresponding to the final
state mass range $1.1 < W < 1.6\GeV$.
In terms of $\xi$, the integration of the structure function for
each $Q^2$ corresponds to the area under the resonance curve from
$\xi_{\rm min}=\xi(Q^2,W=1.6\GeV)$ to
$\xi_{\rm max}=\xi(Q^2,W=1.1\GeV)$.
The contribution from the remaining $\xi$ interval,
$0 < \xi < \xi_{\rm min}$, corresponds to the higher $W$ region.
For this we assume that the structure functions are given by the
leading twist contributions, calculated from the MRST parametrization
\cite{MRST}.

In Fig.~\ref{fig:AdlerSR} the individual contributions from the QE,
resonance, and DIS regions are plotted as a function of $Q^2$.
The (positive) QE contribution is large at low $Q^2$ but falls
rapidly with increasing $Q^2$.
The resonant piece of the sum is negative, and partially cancels
the QE component.
The deep inelastic component grows with $Q^2$, since $\xi \to 1$
as $Q^2 \to \infty$, and for $Q^2 > 1\GeV^2$ contributes some 80\%
of the integral.
Combining the three terms, the sum rule is found to be satisfied
within $\sim 10\%$ over the whole range $0.5 < Q^2 < 2 \GeV^2$.

Since the Adler some rule is based on very general grounds, one
expects it to be exact.
The 10\% deviation of the calculated sum rule from the exact value
should therefore be treated as an indication of the accuracy of the
model.
In practice, the uncertainty comes mainly from the axial form
factors for the second resonance region, and suggests that some
of them are underestimated.
The requirement that the Adler sum rule is satisfied exactly
could therefore serve as a restriction on the currently unknown
axial form factors.

\section{Conclusion \label{Conclusions}}

Motivated by the need to understand neutrino--nucleon cross sections in
the $Q^2 \sim$~few GeV$^2$ range, and the observation of quark-hadron
duality in electron--nucleon scattering, we have performed a detailed
phenomenological study of duality in neutrino structure functions.
Using a recently developed model \cite{Pas03,Lal05,Lal06} for the first
four lowest-lying nucleon resonances, we have computed the structure
functions $F_2$, $2xF_1$ and $xF_3$ in the resonance region for
proton and neutron targets, and compared these with leading twist
parameterizations.

As a check of the resonance model, we have calculated the
electron--nucleon structure functions and found that for each
resonance these oscillate around the leading twist curves down
to low values of $Q^2$, in qualitative agreement with duality.
For quantitative comparisons, we defined ratios $I_i(Q^2)$ of
resonance to leading twist structure functions, which in the
ideal case of duality, should be unity.
Our results show that for the $F_2^{eN}$ structure function this
ratio is below unity at low $Q^2$, and slowly grows with $Q^2$,
consistent with recent experimental results \cite{Nic00}.
The agreement with duality for $0.5 \lesssim Q^2 \lesssim 2\GeV^2$
in this case is at the level of 20\%.
At low $Q^2$ the resonance averaged $F_2$ structure function resembles
valence quark distributions, apparently oblivious to sea quark effects,
which supports the hypothesis of two--component duality \cite{Har69}.
For the $2x F_1$ structure function the ratio is about 40\% above
unity, but would be reduced after correcting for target mass effects
in the leading twist structure function.

For charged current neutrino scattering, duality does not hold for
proton and neutron targets separately because of the dominant role
played by the isospin-3/2 resonances.
However, averaging over proton and neutron targets leads to large
cancellations between $I=3/2$ and $I=1/2$ resonances, so that duality
holds at the 20\% level for isoscalar  $\nu N$ structure functions.
Furthermore, the ratios $I_2^{\nu N}(Q^2)$ and $I_3^{\nu N}(Q^2)$
appear to reach constant values already for $Q^2 \approx 1 \GeV^2$.

Another interesting feature of our analysis is that the ratios
$I_1^{eN}(Q^2)$ and $I_1^{\nu N}(Q^2)$ of the $2x F_1$ structure
functions are consistently above unity.
This may be an indication of the importance of target mass corrections
in the leading twist $F_1$ structure functions, which are known to be
more important than those in $F_2$.

In these comparisons we have used leading twist structure functions
obtained from global parton distributions, which are well constrained
by experimental data, especially for $F_2$.
For the resonances, on the other hand, the data are very sparse,
and theoretical input needs to be used.
Our results therefore have an inherent uncertainty arising from poor
knowledge of the transition form factors, particularly at high $Q^2$.

The results obtained here raise the following question: what is the
most efficient and quantitative method for comparing the resonance
contributions with the scaling curves and their QCD corrections?
One approach is to compare the various contributions to sum rules,
in which integrals over resonance and DIS contributions must reproduce
physical constants.
To this end we computed the various contributions to the Adler sum
rule as a function of $Q^2$.
This exercise shows how the relative contributions vary with $Q^2$,
and saturate $\sim 90\%$ of the sum rule.
The remaining 10\% could be accounted for by including more resonances,
and by better determining the transition form factors.

Overall, our quantitative study of neutrino reactions indicates that
duality in structure functions, averaged over protons and neutrons,
is expected to work to even better accuracy for neutrino scattering than
for electron scattering.

\section*{Acknowledgements}

Authored by Jefferson Science Associates, LLC under U.S. DOE Contract No.
DE-AC05-06OR23177. The U.S. Government retains a non-exclusive, paid-up,
irrevocable, world-wide license to publish or reproduce this manuscript
for U.S. Government purposes. The financial support of BMBF, Bonn under
contract 05HT 4 PEA/9 is gratefully acknowledged. EAP and OL wishes to
thank Prof.~A.~W.~Thomas for hospitality in the Jefferson Lab Theory
Center, where part of this work has been carried out.


\end{document}